\documentclass[twoside,twocolumn,9pt]{article}  

\usepackage[version=4]{mhchem}
\usepackage{siunitx}
\usepackage{mathptmx}
\usepackage[scaled=0.92]{helvet}  
\usepackage{microtype}
\usepackage[british]{babel}
\usepackage{graphicx}
\usepackage[a4paper, margin=2cm, bottom=3cm]{geometry}
\usepackage[format=hang, labelfont=bf]{caption}
\usepackage[colorlinks, citecolor=black]{hyperref}
\usepackage[format=plain,justification=justified,singlelinecheck=false,font={stretch=1.125,small,sf},labelfont=bf,labelsep=space]{caption}
\usepackage{xcolor}

\usepackage{booktabs} 
\usepackage{rsc}

\DeclareSIUnit\angstrom{\text{\AA}}
\newcommand{\adam}{\ce{C10H16}}

\graphicspath{{figures}}

\newcommand*{\citen}[1]{%
  \begingroup
    \romannumeral-`\x 
    \setcitestyle{numbers}%
    \cite{#1}%
  \endgroup   
}

\title{\sffamily\bfseries Dynamics in the ordered and disordered phases\\ of barocaloric adamantane}
\author{Bernet E. Meijer\textit{$^{a}$} \and
 Richard J. C. Dixey\textit{$^{a}$} \and
 Franz Demmel\textit{$^{b}$} \and
 Robin Perry\textit{$^{c}$} \and
 Helen C. Walker$^{\ast}$\textit{$^{b}$} \and 
 Anthony E. Phillips$^{\ast}$\textit{$^{a}$} } 
\date{}

\begin{document}
\maketitle


\renewcommand{\thefootnote}{\fnsymbol{footnote}}
\footnotetext{$^{a}$~School of Physics and Astronomy, Queen Mary University of London, London E1 4NS, U.K.}
\footnotetext{$^{b}$~ISIS Neutron and Muon Source, Rutherford Appleton Laboratory, Didcot OX11 0QX, U.K.}
\footnotetext{$^{c}$~Department of Physics and Astronomy, University College London, London WC1E 6BT, U.K.}
\footnotetext{\dag~Electronic Supplementary Information (ESI) available. See DOI: 00.0000/00000000.}
\footnotetext{$^{\ast}$~Corresponding authors. E-mail addresses: a.e.phillips@qmul.ac.uk, helen.c.walker@stfc.ac.uk}

\vspace{16pt}



\begin{abstract}\normalsize\noindent
High-entropy order-disorder phase transitions can be used for efficient and eco-friendly barocaloric solid-state cooling. 
%
Here the barocaloric effect is reported in an archetypal plastic crystal, adamantane.
Adamantane has a colossal isothermally reversible entropy change of 106~JK$^{-1}$kg$^{-1}$. Extremely low hysteresis means that this can be accessed at pressure differences less than 200 bar.

Configurational entropy can only account for about $40$\% of the total entropy change; the remainder is due to vibrational effects. 
Using neutron spectroscopy and supercell lattice dynamics calculations, it is found that this vibrational entropy change is mainly caused by softening  in the high-entropy phase of acoustic modes that correspond to molecular rotations.
We attribute this behaviour to the contrast between an `interlocked' state in the low-entropy phase and sphere-like behaviour in the high-entropy phase. 
%
%
Although adamantane is a simple van der Waals solid with near-spherical molecules, this approach can be leveraged for the design of more complex barocaloric molecular crystals.
Moreover, this study shows that supercell lattice dynamics calculations can accurately map the effect of orientational disorder on the phonon spectrum, paving the way for studying the vibrational entropy, thermal conductivity, and other thermodynamic effects in more complex materials.

\end{abstract}

\vspace{24pt}

\section{Introduction}

With cooling and refrigeration accounting for over 25\% of global energy consumption\citep{UNEP2018} and 10\% of emissions\citep{UBBEI2016}, it has become clear that innovation in this area is a key factor for reaching our climate goals. 
The emissions from prevalent vapour-compression technology are both direct and indirect: direct emissions come from the refrigerants, that are greenhouse gases commonly thousands of times more potent than CO\textsubscript{2}; indirect emissions originate from the electricity use, and unfortunately the efficiency of vapour-compression technology is plateauing \citep{APECEnergyWorkingGroup2018}.
Improvements in sustainable cooling will therefore lie in alternative technologies. Currently, one of the most promising is solid-state cooling using caloric materials: these materials do not cause direct emissions and offer the potential of increased cooling efficiency.

The functional behaviour of caloric materials relies on phase transitions with large entropy changes, induced by an external field. The group of barocalorics, which undergo a pressure-induced phase transition, is especially promising. These materials are abundant and cost-effective, and since they are pressure-driven, their deployment does not require a complicated refrigeration design. 
The remaining challenge in this field is to identify materials that can beat the efficiency of current vapour-compression technology, and that together give us a broad range of operating temperatures that is needed for widespread deployment.

The structure-space for barocalorics is vast, ranging from framework materials to shape memory alloys \citep{Boldrin2021}. A particularly promising group are the orientationally-disordered (or `plastic') crystals, which have shown giant and colossal barocaloric effects \citep{Lloveras2015, Lloveras2019, Li2019, Aznar2021, Aznar2020, Zhang2022}, and it is hoped that this group might host many more efficient barocalorics. 
Since the efficiency is proportional to the entropy change over the phase transition, the search for these materials must be focused on entropy as a design principle. 
However, most studies only consider one type of entropy contribution\citep{Butler2016} and therefore fail to provide a complete picture of the entropy change. 
A first step towards finding the most efficient barocalorics is to unravel all contributions to the entropy, their importance in plastic crystals, and their corresponding molecular origins.

Here, we study the barocaloric effect and its microsopic origins in the plastic crystal adamantane.
%
%
We chose this material for two reasons. 
First, adamantane is an archetypical example of a crystal that is both orientationally disordered and literally plastic, with a waxy consistency and high susceptibility to external stress. For these reasons, it seems highly likely to be a barocaloric; however, despite recent reports of barocaloric behaviour in adamantane derivatives\citep{Aznar2021}, to our knowledge this has not previously been reported in adamantane itself.
%
Second, adamantane's physical properties make it an ideal model system for plastic crystals: it has rigid, near-spherical molecules and its intermolecular interactions are dominated by van der Waals dispersion forces.
This is both encouraging for generalising our results to the wider family of molecular barocaloric materials, and practically useful since its simplicity makes it a good test case for the analysis of vibrational entropy that we develop here.
%

In this work, we reveal adamantane's barocaloric effect through calorimetric measurements. The effect can be classed as `colossal' (following barocaloric terminology\citep{Boldrin2021, Aznar2021, Zhang2022, Li2021}) and we predict it can be accessed with full reversibility under pressures as low as 200 bar. 
We show that more than half of adamantane's large entropy change can be attributed to vibrational effects, and therefore set out to uncover the microscopic mechanisms that give rise to the vibrational entropy change.
%
We do this by performing supercell lattice dynamics calculations followed by band unfolding\citep{Overy2017, Overy2016}, here for the first time applied to an orientationally disordered supercell. 
The model is validated with single-crystal neutron spectroscopy and further supported by a quasi-elastic neutron scattering experiment under high pressure.
%
%
In the high-temperature phase, the acoustic modes soften and are associated with rolling molecules, rather than translations.
This behaviour can be attributed to the change from an interlocking structure at low temperatures to a spherical close-packed structure at high temperatures. 
The principal dynamical mechanism of this archetypical plastic crystal is thus revealed, which can be leveraged in future barocaloric design.

%
%


\section{Background}

\begin{figure}
    \centering
    \includegraphics[width=\linewidth]{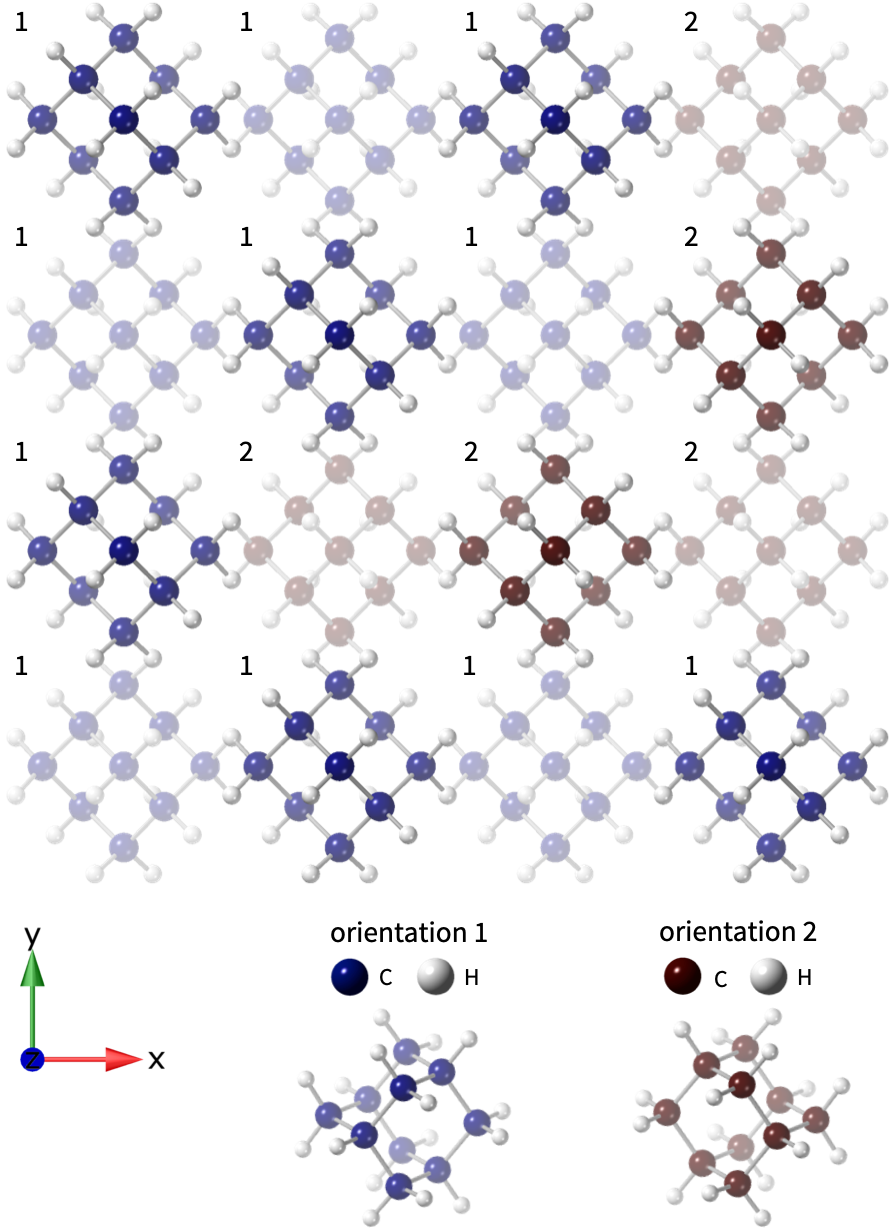}
    \caption{Relaxed configuration of a disordered supercell of adamantane. The two different orientations are highlighted by the numbers and differing colours; the depth fading shows two adjacent planes of the close-packed structure. This is a small part of an $8\times 8\times 8$ supercell used in the calculations, in which the two orientations are present in equal amounts.}
    \label{fig:img_supercell}
\end{figure}
Adamantane is a stable hydrocarbon with formula \adam, which consists of a rigid tetrahedron of six-membered carbon rings in the ``armchair'' configuration (see figure \ref{fig:img_supercell}).  
%
At room temperature and atmospheric pressure, the crystal structure is face centred cubic ($Fm\bar{3}m$, $a=9.426$~\AA, $Z=4$)\citep{Fort1964, Nordman1965}. In this phase, the molecules are randomly oriented in one of two orientations \citep{Amoureux1980, Beake2017}, and they can jump between them via rotation about the fourfold axes, as revealed by NMR\citep{Amoureux1980a} and quasi-elastic neutron scattering studies \citep{Bee1980}.
Upon cooling to $T=208$~K at ambient pressure, 
adamantane undergoes a first-order structural phase transition to the tetragonal space group $P\bar{4}2_1c$ ($a=6.614$~\AA, $c=8.875$~\AA, $Z=2$)\citep{Nordman1965}. In this phase there is no orientational disorder: the planes of molecules alternate down the tetragonal $c$ axis between the two high-temperature orientations.

\begin{table}[h]
\small
    \begin{tabular*}{0.48\textwidth}{@{\extracolsep{\fill}}llr}
    \toprule
    {\textbf{Quantity}} & {\textbf{Value}} &  {\textbf{Reference}} \\
    \midrule
    $T_t$ (ambient $p$) & 208 K & \citen{Nordman1965} \\
    $p_t$ (ambient $T$) & 4.8 kbar & \citen{Breitling1971, Ito1973, Vijayakumar2001} \\
    $\Delta S$ & 16.2 J\,K$^{-1}$\,mol$^{-1}$ & \citen{Chang1960} \\
    $\Delta H$ & 3.38 kJ\,mol$^{-1}$ & \citen{Chang1960} \\
    $\Delta V_{\text{LT}\rightarrow\text{HT}}/V_{\text{HT}}$ & 7.34\% & \citen{Nordman1965}\\
    \bottomrule
    \end{tabular*}
    \caption{Thermodynamic data of adamantane's phase transition. $\Delta S$, $\Delta H$, $\Delta V_{\text{LT}\rightarrow\text{HT}}$ are the thermodynamic changes of the temperature-induced phase transition at ambient pressure. $\Delta V_{\text{LT}\rightarrow\text{HT}}V_{\text{HT}}$ is the volume change of the unit cell from the low-temperature (LT) to high-temperature (HT) phase as a percentage of the high-temperature unit cell volume.}
    \label{tab:thermo}
\end{table}

The thermodynamic data of adamantane's phase transition are summarised in table \ref{tab:thermo}.
As is hinted by the large volume change, the phase transition can also be induced by pressure: at ambient temperature, this happens at a pressure of $p=4.8$~kbar\citep{Breitling1971, Ito1973, Vijayakumar2001}. Both the large volume change and the phase transition temperature's strong sensitivity to pressure are indicators of potentially large barocaloric effects. 

On top of the quasi-elastic neutron studies, inelastic neutron studies have been performed in adamantane's plastic phase\citep{Windsor1978, Windsor1981, Damien1978}. Here we expand upon the quasi-elastic studies by measuring the reorientational dynamics under pressure; the inelastic studies are extended by probing the dynamics in \textit{both} phases to reveal the phase transition mechanism.

\section{Experiment}

\subsection{Colossal barocaloric effect}\label{sec:bar_effect}

\begin{figure*}[h]
    \centering
    \includegraphics[width=\linewidth]{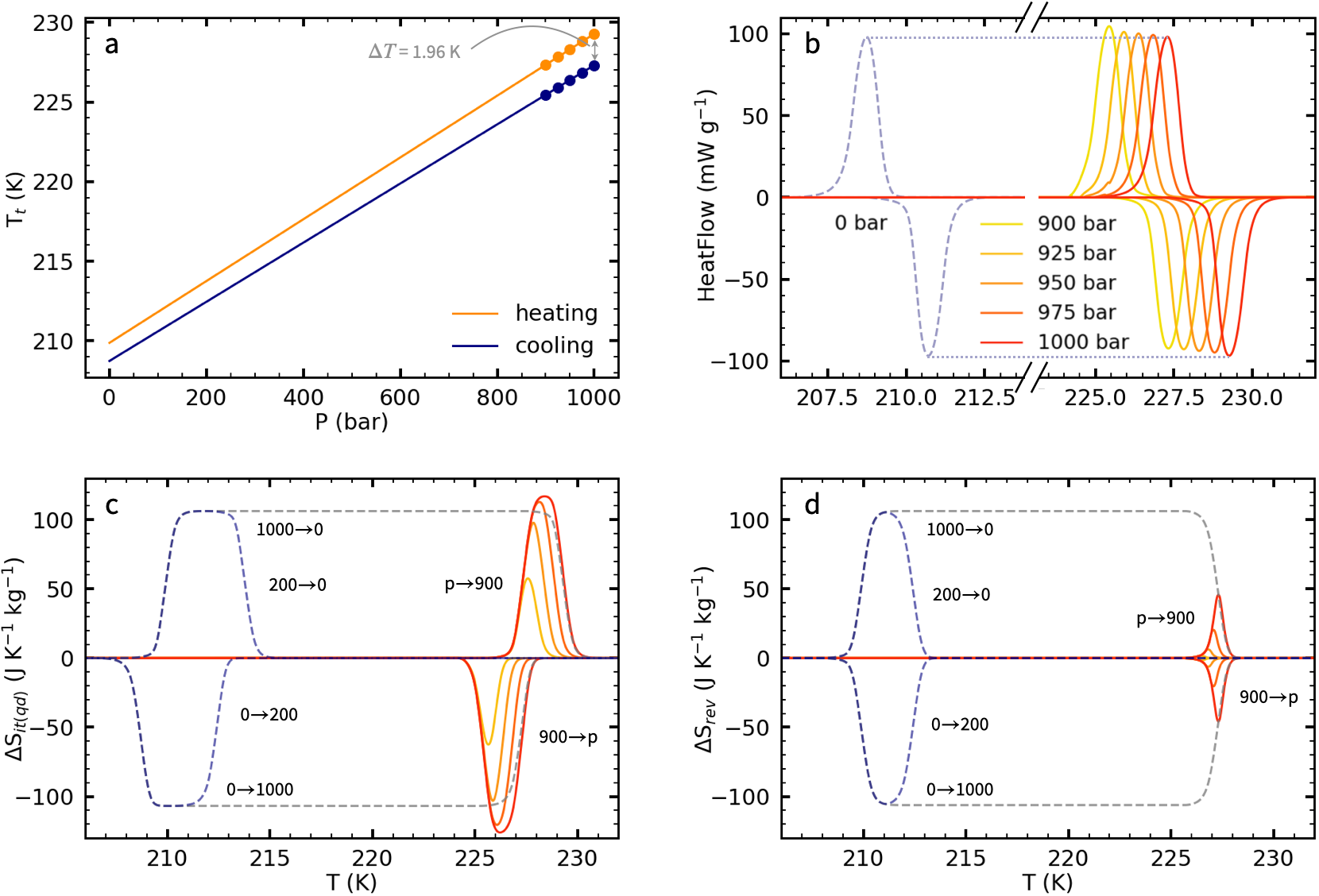}
    \caption{Colossal barocaloric effects in adamantane. (a) The phase diagram as deduced from heat-flow measurements shows a strong pressure-induced shift in the phase transition temperature ($dT_t/dp = 18.57$~Kkbar$^{-1}$ on cooling and $dT_t/dp = 19.40$~Kkbar$^{-1}$ on heating), and extremely small hysteresis up to 1.96 K at 1000 bar. The line shows a linear fit to the data extrapolated down to 0 bar. (b) Heat flow data: the solid lines are experimental data and the dotted line is extrapolated data of 0 bar following the procedure described in the main text. (c) Integrating under the heat flow peaks gives the quasi-direct isothermal entropy change $\Delta S_{\text{it(qd)}}$ for $900$ bar $\rightarrow p$ and $p \rightarrow 900$ bar. The dotted lines are extrapolated entropy change: the navy dotted line is for $0\leftrightarrow200$ bar and the grey dotted line is for $0\leftrightarrow1000$ bar. (d) Reversible entropy change $\Delta S_{\text{rev}}$ deduced from (c).}
    \label{fig:calorimetry}
\end{figure*}


The barocaloric properties of adamantane were measured using high-pressure differential scanning calorimetry (DSC). Figure \ref{fig:calorimetry}(a) and (b) show the heat flow measurements and the corresponding phase diagram. The heat flow measurements were performed at pressures between 900 and 1000 bar. At lower pressures, the phase transition temperature is below the temperature range of the DSC so it could not be observed. 
The phase diagram confirms the strong pressure-induced shift of the phase transition temperature $T_t$, with $dT_t/dp = 18.57$~K\,kbar$^{-1}$ on cooling and $dT_t/dp = 19.40$~K\,kbar$^{-1}$ on heating. It also reveals an extremely low hysteresis up to 1.96 K at 1000 bar, which varies very slightly with pressure. By subtracting the integral of the heat flow peaks at $p>900$ bar from the integral of the $p_{\text{base}}=900$ bar peak, we can recover the isothermal entropy change for releasing pressure down to 900 bar ($p\rightarrow 900$ bar) and adding pressure starting from 900 bar ($900$ bar $\rightarrow p$). This is shown in figure \ref{fig:calorimetry}(c). The maximum pressure-induced entropy change in this experiment is 116.92~J\,K$^{-1}$\,kg$^{-1}$.
Finally, the reversible entropy changes are shown in figure \ref{fig:calorimetry}(d). Starting at a pressure of 900 bar, reversible effects are already achieved with a pressure of 50 bar to reach 950 bar.

To get an estimate of the barocaloric behaviour at low temperatures and pressures, and to find the pressure at which full reversibility is reached, the heat flow data were extrapolated down to 0 bar. First, the phase transition temperature was extrapolated down to 0 bar using a linear fit to the data shown in figure \ref{fig:calorimetry}(a). Next, the 1000 bar heat flow peak was translated to those phase transition temperatures: as an example, the resulting predicted heat flow data for 0 bar are shown in figure \ref{fig:calorimetry}(b). Using the predicted 0 bar data, the isothermal entropy change for $p\leftrightarrow 0$ bar can be estimated; the prediction for $1000\leftrightarrow 0$ bar is shown in figure \ref{fig:calorimetry}(c), along with the prediction for $200\leftrightarrow 0$ bar (which uses the two extrapolated heatflow datasets of 0 and 200 bar). Further details of behaviour at intermediate pressures is available in the supplementary material. 
Finally, this results in the extrapolated reversible entropy changes shown in figure \ref{fig:calorimetry}(d). 
At $<200$ bar full saturation is thus expected, yielding a colossal reversible entropy change of 106~J\,K$^{-1}$\,kg$^{-1}$. 
%
The small operating pressures that are necessary for adamantane's barocaloric exploitation are very appealing, and are likely a consequence of the extremely small hysteresis. With the reasonable assumption that the $dT_t/dP$ relation is linear, the hysteresis at 0 bar is estimated to have a value of only about 1.15 K, far smaller than that of some adamantane derivatives\citep{Aznar2021}.

\begin{figure}
    \centering
    \includegraphics[width=\linewidth]{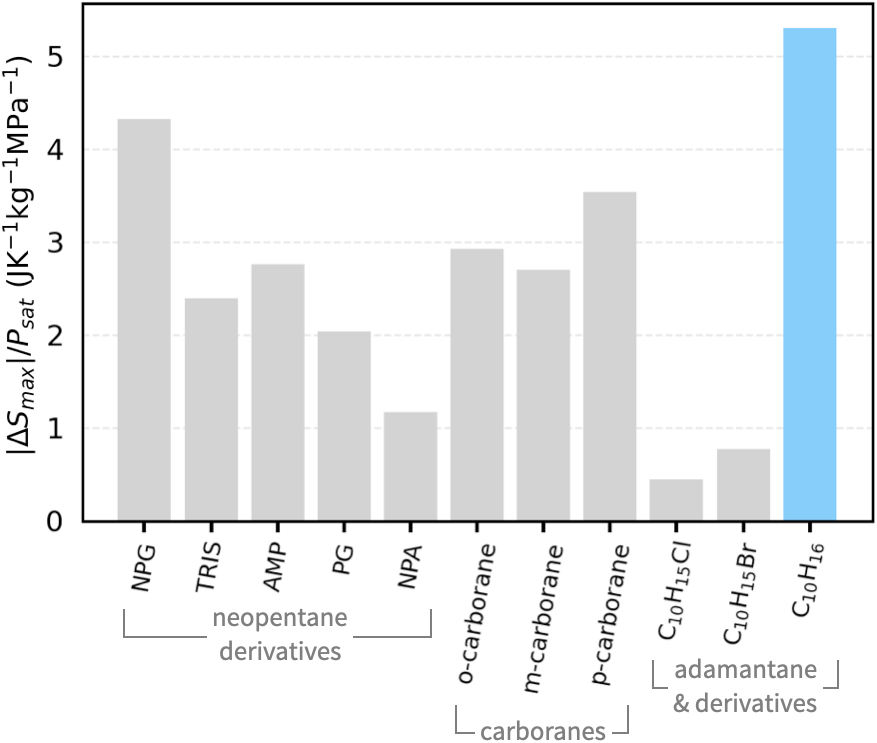}
    \caption{Maximum pressure-induced entropy change $|\Delta S_{\text{max}}|$ normalised by saturation pressure $P_{\text{sat}}$ for various barocaloric plastic crystals, including neopentane derivatives\citep{Li2019, Lloveras2019, Aznar2020}, carboranes\citep{Zhang2022}, adamantane derivatives\citep{Aznar2021} and adamantane as presented in this work (highlighted in blue).}
    \label{fig:baro_comparison}
\end{figure}

The barocaloric properties of adamantane compare favourably against other caloric materials. The isothermal entropy change far surpasses that of electrocaloric and magnetocaloric materials (with maximum entropy changes not greater than $\sim50$~JK$^{-1}$kg$^{-1}$\citep{Boldrin2021, Li2019}). Moreover, the reversible entropy change is also among some of the largest observed in barocaloric plastic crystals (carboranes: $72-97$~JK$^{-1}$kg$^{-1}$\citep{Zhang2022}; 1-X-adamantanes: $\sim150$~JK$^{-1}$kg$^{-1}$ (at 1 kbar)\citep{Aznar2021}; NPG: $389$~JK$^{-1}$kg$^{-1}$\citep{Li2019, Lloveras2019}) and crucially, due to the phase transition temperature's high sensitivity on pressure, adamantane's maximum entropy change normalised by saturation pressure is larger than any barocaloric plastic crystal known so far, as shown in figure \ref{fig:baro_comparison}. This means that entropy changes can be achieved with minimal work. Although adamantane's low phase transition temperature makes it unsuitable for most domestic applications, it may be an excellent candidate for ultra-low temperature freezers used in vaccine storage\citep{U.S.DepartmentofHealthandHumanServices-CentersforDiseaseControlandPrevention2022} (e.g. the Moderna\citep{Moderna2022} and Johnson\&Johnson\citep{Johnson2022} COVID-19 vaccines), blood banks\citep{Shabihkhani2014} and forensic labs\citep{Dash2021}.


\subsection{Probing phonons with neutron scattering}

Dispersion curves for adamantane's low- and high-temperature phases were collected with a single-crystal, inelastic neutron scattering experiment and are shown in figures \ref{fig:phonons}(a) and \ref{fig:phonons}(c) respectively. The low-temperature tetragonal phase has been plotted in quasi-cubic coordinates to allow for direct comparison to the cubic high-temperature structure. Data in this phase has also been collected at $T=220$~K by the application of pressure and the dispersion curve is available in the supplementary material. These two dispersion curves show only small differences, which can be attributed to slightly different crystal twinning.

In the high-temperature phase there is a substantial amount of phonon broadening, which can in principle be attributed to either disorder or anharmonicity. Nevertheless, the cell doubling between the two phases is clearly visible (as evidenced by the acoustic modes out of the X-point in the low-temperature phase), and the acoustic modes seem to soften in the high-temperature phase, especially in the $\Gamma$--$L$ direction, which is the close-packing direction. Details of the softening of the acoustic modes and phonon density of states are available in the supplementary material. Softening of acoustic modes is characteristic of the plastic phase, in which the elasticity decreases. 

\subsection{Reorientational dynamics under pressure}

Previous studies have shown that in the high-temperature phase, adamantane molecules dynamically reorient between their two orientations with fourfold jumps\citep{Amoureux1980a, Bee1980}. A recently developed high-pressure gas cell has made it possible to investigate this dynamical behaviour under pressure, i.e. under working conditions of a barocaloric cooling cycle.
It was found that the reorientational dynamics are suppressed by pressure. When applying a pressure of 4 kbar, the average time between jumps increases by a factor of $2.7$. Not only the frequency, but also the fraction of dynamically activated molecules decreases with pressure: with 4 kbar, this fraction decreased by a factor of $0.6$. Details of the results and data analysis are available in supplementary material.

\section{Modelling}
\subsection{Phonons in the ordered and disordered phases}\label{sec:gdos}

\begin{figure*}[!htpb]
    \centering
    \includegraphics[width=\linewidth]{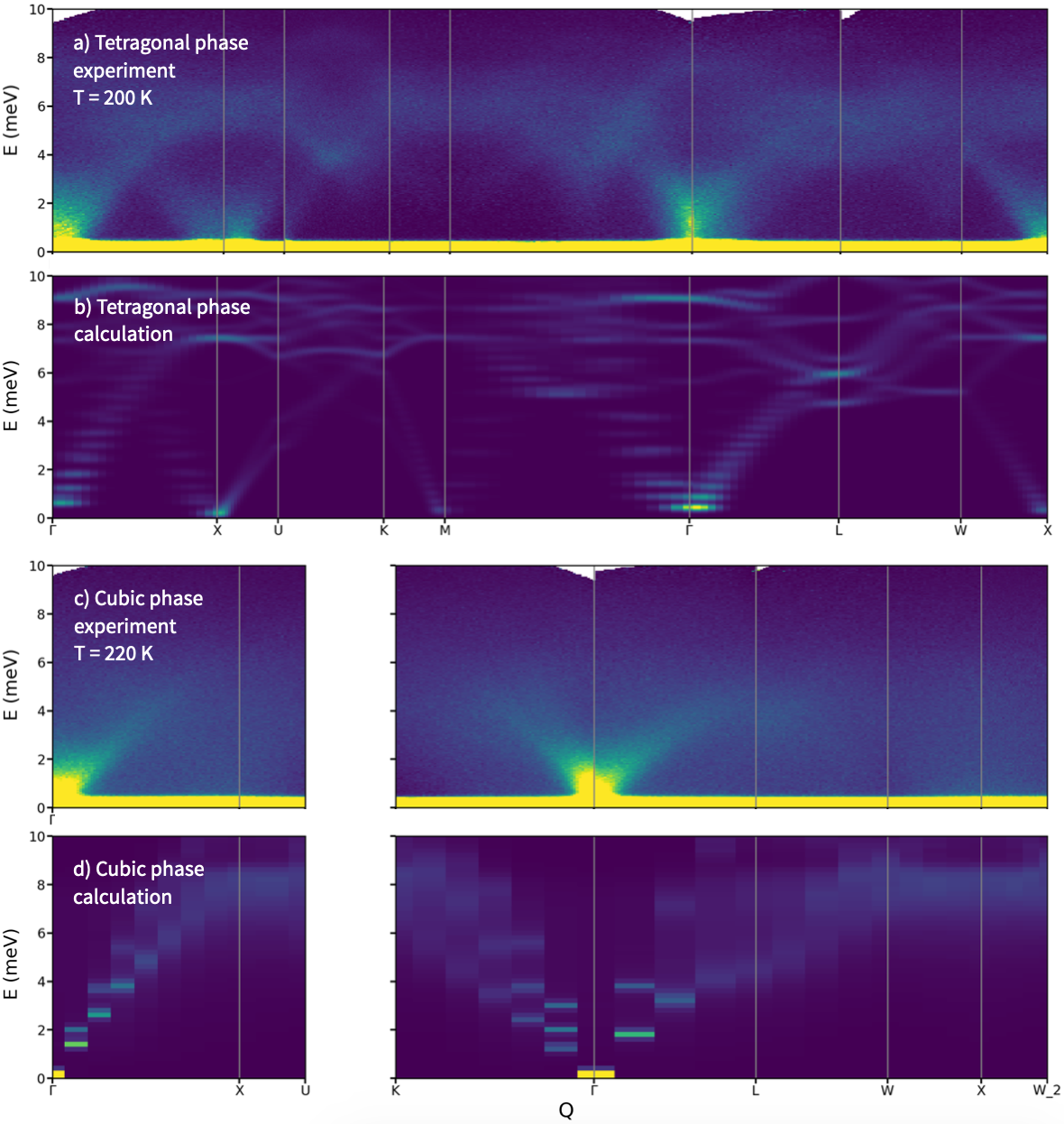}
    \caption{Phonon dispersion of the tetragonal low-temperature phase and cubic high-temperature phase of adamantane. (a) and (b) are the measured and calculated disperions for the low-temperature tetragonal phase respectively; idem for (c) and (d) for the high-temperature cubic phase. The calculated dispersion in (b) is neutron-weighted and there has been accounted for crystal twinning by taking a weighted average of the dispersions of the three crystal twins. The high-temperature phase is \textit{not} neutron-weighted because it has been calculated with the SCLD approach. An $8\times8\times8$ cubic supercell was used, in which each molecule was randomly assigned one of the two possible orientations.}
    \label{fig:phonons}
\end{figure*}

Experimentally, the entropy change between the phases of adamantane is 106~JK$^{-1}$kg$^{-1}$. However, the Boltzmann formula for configurational entropy predicts an entropy change of only $(\ln 2) R = \SI{42.30}{J.K^{-1}.kg^{-1}}$  (or $\SI{5.76}{J.K^{-1}.mol^{-1}}$) due to the twofold configurational disorder in the high-temperature phase. 
A large part (about 60\%) of the total entropy change must therefore originate from vibrational effects.
To investigate the molecular mechanism that gives rise to the increase in vibrational entropy over the phase transition, we perform supercell lattice dynamics calculations using an empirical potential, validated against experimental dispersion curves.



The vibrations in adamantane were modelled using lattice dynamics calculations with the program GULP\citep{Gale2003}, using the forcefield parameterised by \citet{Greig1996}. In this model, the adamantane molecules are taken as rigid, and the intermolecular forces are parameterised by a Buckingham potential. This model has been shown to successfully predict both the ordered and disordered structure\citep{Greig1996} and also reproduces the pressure-induces phase transition to the ordered phase\citep{Murugan2005}. The model could also be formulated with flexible rather than rigid molecules, but this has not shown to have a substantial effect on the structural properties below 50~kbar\citep{Murugan2005a}. 

The phonon spectrum in the low-temperature phase, calculated using standard lattice dynamics calculations, is shown in figure \ref{fig:phonons}(b) and shows good agreement with the measured dispersion. All features in the dispersion curve are reproduced, albeit with slightly higher energies in the calculation. This, however, is only a small discrepancy given the use of a classical forcefield. The accuracy of the vibrational energies might potentially be increased by further fitting of the forcefield. However, as mentioned above, the model reliably predicts adamantane's structure\citep{Greig1996, Murugan2005}, and since all qualitative dispersion features are reproduced, we did not deem this necessary. We note that the good match of the experimental dispersion curve to our rigid-molecule model implies that internal vibrations of the adamantane molecules occur at high energies not accessible with this experiment. These modes, in turn, will therefore have a minor effect on the vibrational entropy change, as is clear from equation \ref{eq:S_vib} below.

In the high-temperature phase, a disordered supercell configuration is required to accurately capture the effect of the configurational disorder on the phonon dispersion relation. We use the so-called Supercell Lattice Dynamics (SCLD)\citep{Overy2016, Overy2017} approach: phonons are calculated in a disordered supercell at the $\Gamma$ point and consequently unfolded over the first Brillouin zone of the unit cell. The resulting phonon dispersion can be directly compared to dispersion relations measured by experiment. To date, the SCLD approach has been used to model disorder-induced phonon broadening for mass- and force-constant disorder; here, we extend this approach to orientational disorder.
Figure \ref{fig:phonons}(d) shows the calculated dispersion relation in the high-temperature phase: the $\Gamma$-point phonons were calculated for an $8\times8\times8$ cubic supercell, in which each molecule was randomly assigned one of the two allowed orientations; the modes were consequently unfolded over the first Brillouin zone of the cubic unit cell.
The calculation has imaginary frequencies at the $\Gamma$ point (see supplementary material for a dispersion plot). With the methodology explained in the next section, we were able to identify these dynamical instabilities as rotational modes of the adamantane molecules. 
The instability arises from the fact that the energy optimisation in this structure came close to but never reached a global minimum, despite the use of multiple optimisation strategies (see \textit{Methods}). 
We suspect that this is due to the flatness of the energy landscape in the direction of molecular rotations, due to the weak Van der Waals intermolecular forces and the molecules' near-spherical shape. 
In such a flat energy landscape, harmonic molecular rotations, if not unstable, would have very low energies.
%
%

The SCLD calculation reproduces a very substantial amount of the phonon broadening seen in experiment.
This indicates that the phonon broadening in adamantane's high-temperature phase can for a large part be attributed to orientational disorder rather than anharmonic effects.
The features that remain visible despite the broadening are reproduced by experiment, such as the acoustic mode extending to the $L$ point at around 4 meV. (For additional comparison, cuts at specific $Q$-points are available in the supplementary material.)

As was seen in experiment, the calculated acoustic modes soften in the high-temperature phase. It is these modes specifically that seem to be responsible for most of the vibrational entropy change in this material. In the high-temperature limit, the vibrational entropy change between two harmonic phases $\alpha$ and $\beta$ is given by\citep{Fultz2010}
\begin{align} \label{eq:S_vib}
    \Delta S_{\text{vib}}^{\beta-\alpha} = 3k_B \int_0^{\infty} (g^{\alpha}(\epsilon) - g^{\beta}(\epsilon)) \ln (\epsilon) d\epsilon
\end{align}
where $k_B$ is the Boltzmann constant and $g(\epsilon)$ is the harmonic phonon density of states for energy $\epsilon$. This equation effectively means that low-energy phonon modes contribute more to the entropy change; therefore, the softened acoustic modes in the high-temperature phase are partly responsible for the entropy increase. From the density of states (see supplementary figure S6), the calculated vibrational entropy is 106~J\,K$^{-1}$kg$^{-1}$. This is higher than the predicted excess of $\sim 64$~JK$^{-1}$kg$^{-1}$, but as good as an agreement as can be expected with this empirical forcefield.


\subsection{Investigating the high-entropy phonon modes}\label{sec:rot_dynamics}

\begin{figure*}[h]
    \centering
    \includegraphics[width=\textwidth]{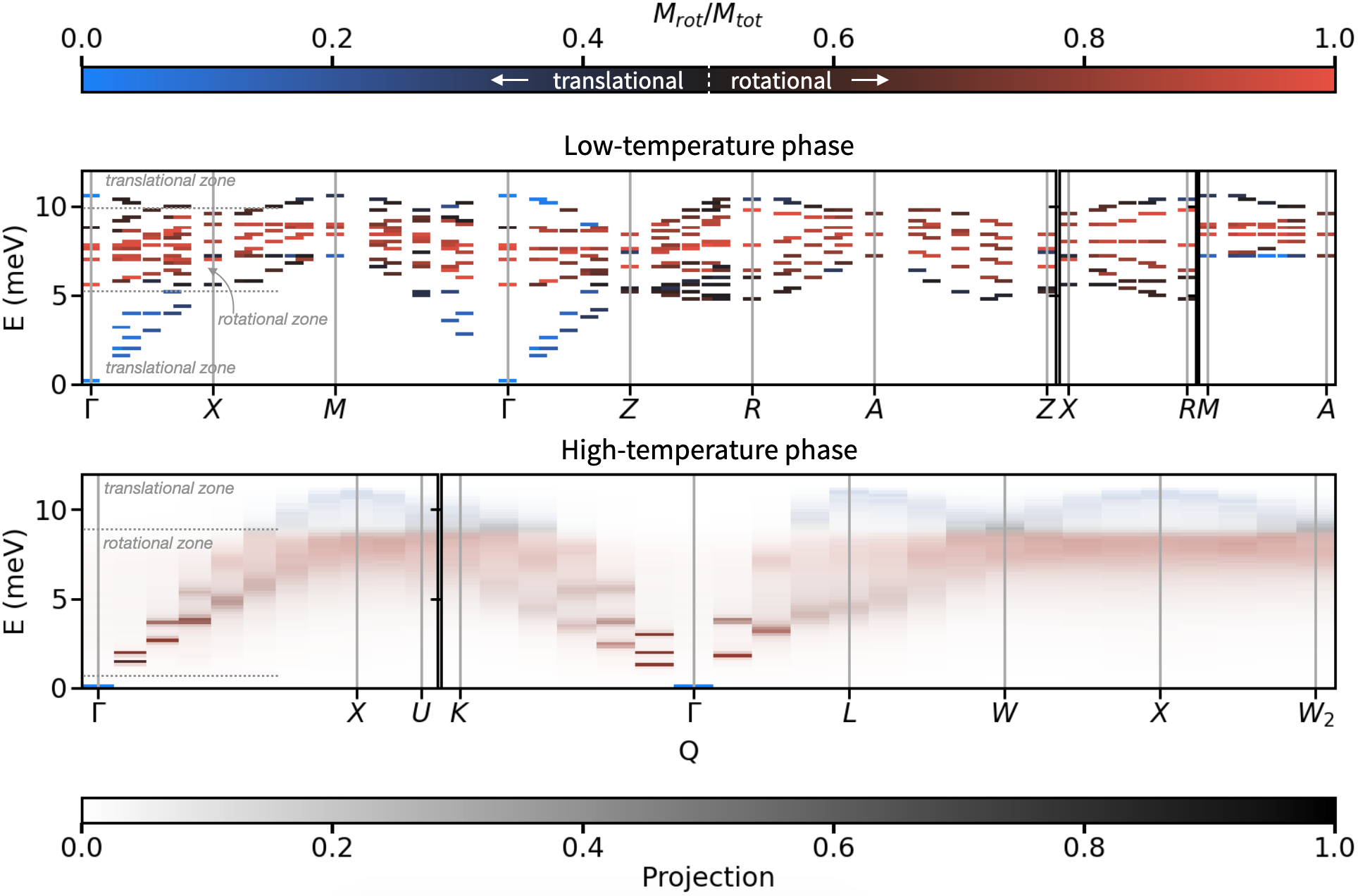}
    \caption{Characterising the phonon dispersion in adamantane's low-and high-temperature phases. For each phase, the modes in the phonon dispersion are coloured by the rotational character of the corresponding eigenvectors, calculated with the SCLD method ($8\times 8\times 8$ supercell in high-temperature phase; $6\times 6\times 6$ and $8\times 8\times 8$ combined in low-temperature phase). The rotational character is the absolute average displacement due to molecular rotation divided by the total absolute average displacement: $M_{\text{rot}}/M_{\text{tot}}$. The rotational character of degenerate modes is averaged. In the high-temperature phase, projected intensity after band-unfolding determines the opacity of the colours (bottom colour bar): modes with a projected value $>1$ have maximum opacity; modes with a projected value of 0 are transparent. The energy spectra show clear `zones' of rotational or translational modes: the dashed lines on the left part of the plots are guides to the eye.}
    \label{fig:character_modes}
\end{figure*}

The softening of the modes in the high-temperature phase -- the root of the vibrational entropy change -- can be investigated further by analysing their character. With the program GASP\citep{Wells2002, Wells2015} we have decomposed the displacements caused by the mode eigenvectors into translations and rotations of the adamantane molecules. Since we consider a rigid-molecule model, molecular distortions are ignored. Each mode in the dispersion relation can then be characterised by its rotational character $M_{\text{rot}}/M_{\text{tot}}$, which is the absolute average displacement due to rotation divided by the total absolute average displacement. The modes are coloured by their rotational character in the dispersion curves in figure \ref{fig:character_modes}.

In the low-temperature phase, we observe a band of rotational modes between approximately 5 and 10 meV. In the high-temperature phase, modes with significant rotational character occur at all energies below 9 meV. Notably, the acoustic modes, which were purely translational in the low-temperature phase, now have a large rotational character. 
As mentioned in the previous section, the few unstable modes are purely rotational (see supplementary figure S4). 

\begin{figure}
    \centering
    \includegraphics[width=\linewidth]{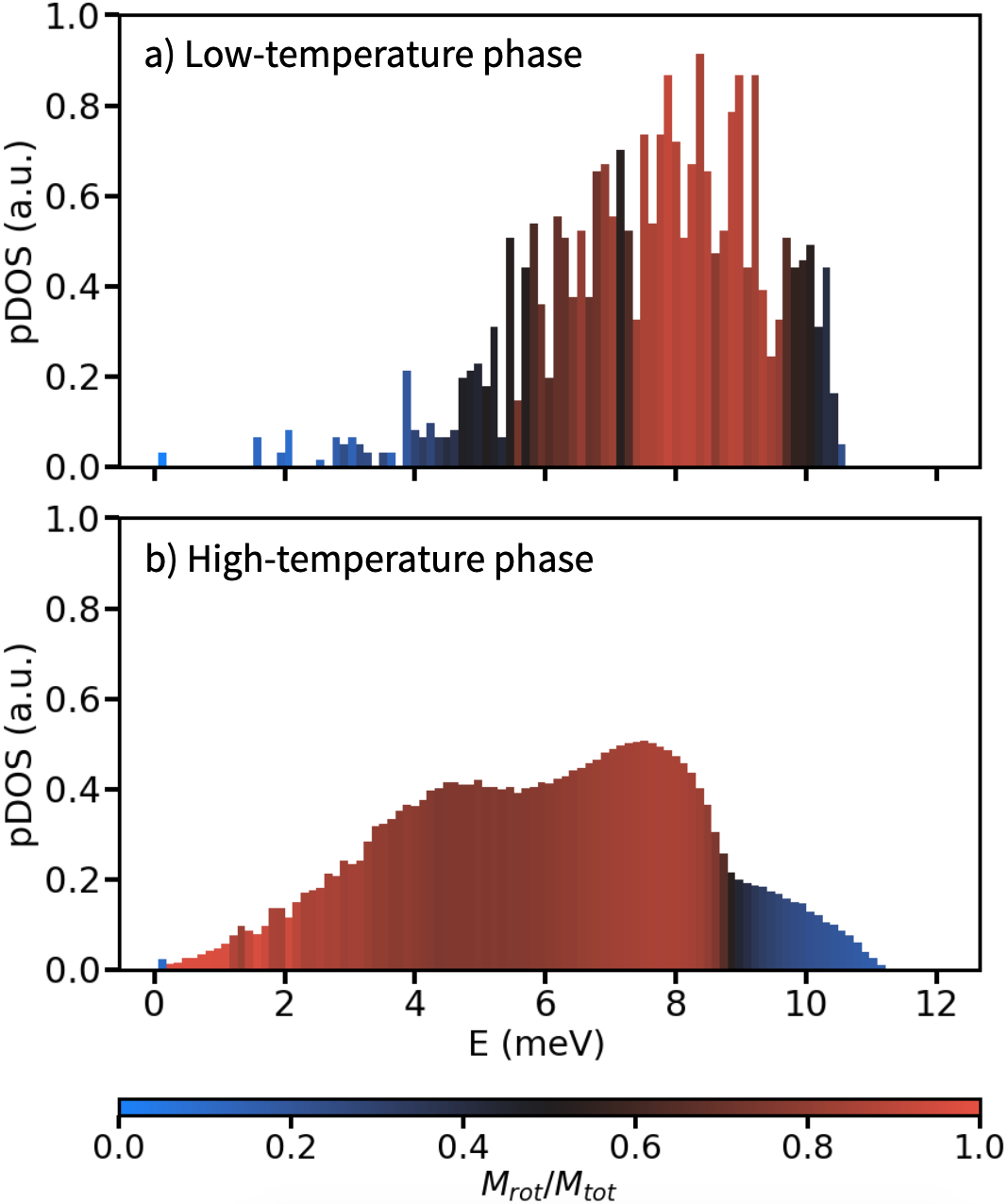}
    \caption{Computational phonon density of states in the high- and low-temperature phases. Each bar in the histograms is coloured by the average rotational character $M_{\text{rot}}/M_{\text{tot}}$ of the modes in that bin.}
    \label{fig:rot_dos}
\end{figure}

The dynamical changes between the two phases are also apparent in the phonon density of states, shown in figure \ref{fig:rot_dos}. Each bin in the phonon density of states histogram is coloured by the average rotational character of the modes in that bin. We clearly see that the low-energy translational modes in the low-temperature phase are replaced by strong rotational modes in the high-temperature phase. 

A possible way of understanding this dynamical change is by contrasting the structures of the low- and high-temperature phase as interlocking molecules and spherical close-packed molecules, respectively.
In the low-temperature tetragonal phase, interlocking decreases the molecules' ability to rotate and hence rotational modes occur at higher energies. At low energies, only translational modes can be excited.
In the high-temperature phase, the structure changes to close-packed FCC. The weak intermolecular interactions between individual atoms no longer play an important role. Instead, the molecules behave as close-packed spheres. In this arrangement, the purely rotational degrees of freedom are the easiest to excite, since rotational motion of one molecule requires barely any cooperation from neighbouring molecules. These purely rotational modes occur at very low energies (and in some cases are unstable in our calculation). In the acoustic modes, the translational and rotational modes start to mix. Away from the $\Gamma$-point, the molecules' ability to translate is inhibited by their close packing, and hence the acoustic modes consist of a rolling behaviour, which is partly rotational and partly translational. The purely translational modes now only occur at higher energies, above 9 meV.

The ability of the molecules to either interlock (low temperature) or behave more like spheres (high temperature) is consistent with adamantane's reorientational dynamics observed by QENS and NMR. 
The existence of reorientational jumps in the high-temperature phase (and their absence in the low-temperature phase) implies that the rotational degrees of freedom are indeed easier to excite in the high-temperature phase. 
The fact that the reorientations in the high-temperature phase are suppressed by pressure can also be interpreted in the light of this simple model: as pressure pushes the molecules closer together, their behaviour slowly starts to shift from sphere-like (easy to reorient) to interlocked (hard to reorient).



\section{Discussion and conclusions}

Plastic crystals have shown to be promising candidates for efficient and environmentally-friendly barocaloric cooling. In this work, we have investigated the barocaloric effect of an archetypical plastic crystal, adamantane. 
The large entropy change over the phase transition - a key metric of its barocaloric efficiency - originates from both orientational disorder (40\%) and vibrational effects (60\%). With supercell lattice dynamics calculations, we have been able to trace back the vibrational entropy change to the softening of (mainly) the acoustic modes. During the softening, these modes also obtain a substantial rotational component and so can be described by rolling molecules. We propose that this behaviour can be attributed to a structural change from an `interlocked' to a spherical close-packed state: at low temperature, adamantane molecules form an interlocking structure, while at high temperatures the molecules behave like close-packed spheres, which unlocks the rotational degrees of freedom at low energies. 
This behaviour is an example of a characteristic feature of a plastic crystal: the dynamical disorder in the plastic phase allows the molecules to obtain a higher \textit{apparent} symmetry (here: spherical) than either its molecular symmetry (here: tetrahedral) or even its site symmetry (here: octahedral). 

The SCLD approach to calculate the phonon spectrum has the great benefit that it allows us to trace the vibrational entropy change back to its \textit{microscopic} origins via an eigenvector analysis. This opens up the possibility to use not only configurational but now also vibrational entropy as a design principle. The good match with the experimentally measured phonon dispersion in this paper demonstrates that the SCLD approach can be applied to orientational disorder, and can hence be a powerful tool in studying many plastic crystals. Forcefields for many plastic crystals are readily available in for example the OPLS all-atom force field for organic and ionic liquids\citep{Jorgensen1996, Tsuzuki2009, Sambasivarao2009}, which has been successfully used in some previous plastic crystal studies\citep{Park2020, Li2019}. The SCLD approach can be readily extended to include quasi- and anharmonic effects by calculating the dynamical matrix from molecular dynamics\citep{Kong2011}, albeit at greater computational cost.

This case study of adamantane has brought insight into the specific properties of plastic crystals in which van der Waals forces dominate the intermolecular interactions. Although adamantane has large reversible entropy changes, its phase transition temperature may restrict its use to low-temperature cooling applications. In addition, though, our detailed analysis of the molecular mechanism underlying its entropy change can inform the design of more sophisticated barocaloric plastic crystals, and can set a benchmark for other plastic crystals in which 1) the interactions are determined by more than just van der Waals forces; and/or 2) the molecules deviate from adamantane's near-spherical shape. 

An obvious first step in this regard will be to compare adamantane's barocaloric behaviour against that of adamantane derivatives such as the recently reported 1-haloadamantanes\citep{Aznar2021}: these compounds have similar entropy changes, available at room temperature, although they show more hysteresis than adamantane. This might suggest that the spherical nature of adamantane's molecules promotes the transition to the close-packed high-temperature phase (low phase transition temperature), and might eliminate the need of a metastable state during the structural change (low hysteresis). However, a full analysis of the entropy contributions and dynamic behaviour (for example with the SCLD approach) of these compounds is needed to make definitive conclusions about the influence of molecular symmetry on barocaloric performance.

Finally, the SCLD approach demonstrated in this paper can help map out another material property vital to barocaloric deployment: thermal conductivity. Configurational disorder in barocalorics is a two-edged sword: on the one hand, configurational disorder in the high-temperature phase contributes to the entropy change; on the other hand, configurational disorder reduces phonon lifetimes and, with them, thermal conductivity. 
Good thermal conductivity is a key requirement for a working barocaloric cooling device, and this presents a challenge unique to solid-state cooling. In particular plastic crystals have low thermal conductivity\citep{Krivchikov2015, Lloveras2021}, with adamantane having only $\sim 0.18$~Wm$^{-1}$K$^{-1}$ in the plastic phase\citep{Aznar2021, Wigren1980}. 
Several engineering solutions have been proposed to mitigate this issue (such as mixing the barocaloric with highly conductive materials\citep{Aznar2021, Praveen2018, Lee2008, Mesalhy2005, Krivchikov2012, Krivchikov2015, Vdovichenko2015}), but many open questions remain regarding the quantitative relation between disorder and thermal conductivity, and the specific conditions that might amplify or suppress this relation\citep{DeAngelis2019}. 
By directly modelling the influence of disorder on phonon lifetimes via phonon broadening, the SCLD approach presents an opportunity to further the understanding of the interplay between disorder, configurational entropy, vibrational entropy (via the broadened density of states) and thermal conductivity, and can also perhaps inspire ways of tuning thermal conductivity\citep{Overy2016} in barocalorics.

\section*{Methods}

\subsection*{Sample preparation}

The single crystal was grown using a slow cooling melt method. $1$~g of powdered isotope-enriched adamantane (CDN Isotopes, $99.3\%$-d16\footnote{INS measurements require perdeuterated samples to avoid the huge incoherent scattering cross section of hydrogen which would swamp the phonon signal.}) was sealed in a thin-walled, $1$~cm diameter quartz tube under vacuum. The tube was heated to $300$~$^\circ$C in a conventional box furnace, then cooled at $2$~$^\circ$C per hour, through the melting point at $270$~$^\circ$C to $100$~$^\circ$C. The tube was then left at $100$~$^\circ$C for one month to anneal the crystal to improve grain size and crystallinity. The resultant crystal mass was around $0.95$~g and confirmed to be a single grain by neutron diffraction.

\subsection*{Calorimetry} \label{sec:method_calor}

Calorimetry measurements, used for the determination of the barocaloric effects of adamantane, were performed on a Seteram MicroDSC7 EVO DSC with a microcalvet sensor. High pressure measurements were performed with a pair of hastelloy high pressure DSC cells, with the empty high pressure cell in the reference well to compensate for the heat capacity of the cell. A 47.3 mg sample of polycrystalline adamantane was sealed within a high pressure stainless steel hastelloy DSC cell, and cooled below the phase transition during temperature and pressure stabilisation.  Quasi-direct barocaloric measurements were carried out between 225 to 240 K, under constant pressures between 925 and 1000 bar in 25 bar steps, using nitrogen as a hydrostatic pressure transmitting medium.  A heating and cooling rate of 1 K/min was used for both the cooling and heating processes. Heat flow data after subtracting baseline background were used for calculation of the entropy changes. The phase transition temperature at ambient pressure is below the operating temperature range of this instrument: the lowest temperature we achieved was $-50^{\circ}$ C and the recommended operating limit is $-45^{\circ}$ C. Therefore, the phase transition was only detectable at pressures starting at 900 bar. Below 900 bar the heating curves have been extrapolated to lower temperatures using the high temperature data as a reference, and scaled to account for the increase in phase transition enthalpy at high temperatures. This ensured that the maximum isothermal entropy change is the same at all pressures. Since entropy changes are likely to be suppressed by high pressure, we expect that this method slightly \textit{underestimates} the entropy change at low pressures.

\subsection*{High-pressure neutron spectroscopy} \label{sec:method_ins}

\subsubsection*{Inelastic}

Inelastic neutron scattering measurements were performed on the single crystal as a function of temperature as well as pressure, as it was part of the commissioning of a recently developed low-background, cylindrical clamp cell for inelastic neutron scattering. The cell is made of TAV6, a titanium alloy composed of $6$~wt\% Al, $4$~wt\% V and $90$~wt\% Ti\citep{Kibble2020}, and minimises unwanted absorption and background scattering that often complicate high-pressure inelastic neutron scattering experiments. The single crystal was inserted into the TAV$6$ pressure cell using helium as a pressure transmitting medium, which was mounted in a helium cryostat on the direct geometry time of flight spectrometer LET\citep{Bewley2011} at ISIS. Measurements were performed using rep-rate multiplication allowing the simultaneous collection of data with incident neutron energies of $1.77$, $3.70$ and $12.14$~meV. Data were collected at ambient pressure in the high temperature cubic phase at $T=220$~K, and in the low temperature tetragonal phase at $T=200$~K. $1$~kbar of pressure was then applied at $T=220$~K. The raw data were reduced and corrected for detector efficiency and time independent background using the Mantid software package\citep{Arnold2014}, while the processed data were analysed using the Horace software package\citep{Ewings2016}.

\subsubsection*{Quasielastic}
High-pressure quasielastic neutron scattering experiments were performed on the indirect-geometry, time-of-flight spectrometer OSIRIS at ISIS, UK. This experiment used the same TAV6 pressure cell as was used for the inelastic experiment. The TAV6 alloy produces parasitic Bragg peaks starting at around $Q=2.5$ \AA{}\textsuperscript{-1}. However, they are not visible to this experiment: here the 002 reflection of the pyrolytic graphite analyser of OSIRIS was chosen, giving an energy resolution function with FWHM 25.4 $\mu$eV, dynamic range of $\pm 0.5$ meV and $Q$-range of 0.18 to 1.8 \AA\textsuperscript{-1}. The 7mm diameter of the inner bore of the cylindrical sample cell gives a relatively large sample thickness, which increases the probability of multiple scattering. This was mitigated by diluting the sample with the low incoherent scatterer KBr: approximately 1.0 g of powdered $>99$\% adamantane (purchased from Sigma Aldrich) was diluted with KBr in a 0.3:1 mass ratio.
In addition to the data collection at various pressures, and empty can measurement (for background subtraction) and low-temperature collection (to obtain experimental resolution) were performed in the same experimental setup. Data reduction and background subtraction was done in Mantid\citep{Arnold2014} and further data analysis was done in python using our own code\citep{Meijer_simpleQENS_2021}.

\subsection*{Calculation} \label{sec:method_calc}

\subsubsection*{Energy minimisation}

Before the dynamical matrix was diagonalised, the energy of the system was minimised. In the low-temperature phase, the energy of the unit cell was minimised using the default Newton-Raphson method in GULP with a Broyden-Fletcher-Goldfarb-Shanno (BFGS)\citep{Shanno1970} Hessian updating scheme.

In the high-temperature phase, an $8\times 8 \times 8$ supercell (of FCC unit cells) was chosen, with a random distribution of the two possible orientations of the molecules. Finding a global energy minimum in this system proved difficult. We therefore used a combination of different algorithms to get as close to a global minimum as possible:
\begin{enumerate}
    \item energy minimisation using the conjugate gradients implementation in the large-scale atomic/molecular massively parallel simulator (LAMMPS) code\citep{Thompson2022}. An external pressure of 1 bar was set to allow the simulation box to change size isotropically during the minimisation process. The procedure was stopped when the energy could not be minimised any further; however, the forces do not fulfil the stopping criteria, which means that a global minimum has not been found;
    \item energy minimisation using the conjugate gradients implementation in GULP. The cell vectors are allowed to vary individually. Similarly to the LAMMPS result, the procedure was stopped when the energy could no longer be minimised, although a global minimum was not found.
    \item energy minimisation using the Newton-Raphson method in GULP. A limited-memory BFGS Hessian updating scheme was used, since calculating the full inverse Hessian in this large system ($\sim 53,000$ atoms) was not possible. Again, this procedure finished when the energy could no longer be minimised, without finding a global minimum.
\end{enumerate}
Optimisation of a sample of disordered FCC \textit{unit} cells using a Newton-Raphson method with standard BFGS or conjugate gradients was also unsuccessful, as was evidenced by the existence of imaginary frequencies after the dynamical matrix diagonalisation. 
Nevertheless, by sequentially optimising the supercell with the methods described above, these imaginary frequencies were reduced to $\sim 580$ modes in the supercell, and after projection only represent 6.1\% of all intermolecular modes in the Brillouin zone. In the calculation of the phonon density of states, the absolute value of these modes was used as an approximation of their true energy. As seen in the supplementary material, including the imaginary modes has a minimal effect on the phonon density of states.

\subsubsection*{Lattice dynamics}
Lattice dynamics calculations were performed using the General Utility Lattice Program (GULP)\citep{Gale2003}. The model consists of intermolecular forces between nearest neighbours, which have been parameterised as a Buckingham potential ($U(r) = A e^{-r/\rho} - C/r^6$) by \citet{Greig1996}. The parameter values are given in table \ref{tab:gulp_params}. The molecules itself are rigid, which is approximated in GULP by setting the harmonic intramolecular bend- and stretch force constants to a large value ($k = 100$~eV\AA$^{-2}$). 
\begin{table}[h!]
  \begin{center}
    \caption{Adamantane intermolecular nearest-neighbour Buckingham potential parameters\citep{Greig1996}.}
    \label{tab:gulp_params}
    \begin{tabular}{l|l|r|r}
      \toprule 
      \textbf{Type} & $A$ (kJ\,mol$^{-1}$) & $\rho$ (\AA) & C (kJ\,mol$^{-1}$\,\AA$^{-6}$)\\
      \midrule 
      C-C & 359824.0 & 0.278 & 2374.42\\
      C-H & 38492.8 & 0.278 & 522.67\\
      H-H & 11715.2 & 0.267 & 113.93 \\
      \bottomrule 
    \end{tabular}
  \end{center}
\end{table}

For the low-temperature phase, the dispersion curve of figure \ref{fig:phonons}(b) is calculated using the quasicubic unit cell. The dynamical matrix was diagonalised using GULP, after which the neutron-weighted phonon dispersion was obtained using the python package Euphonic\citep{Fair}. Neutron weighting can alter the visibility of modes in the simulated phonon dispersion due to the varying coherent scattering cross-sections for different elements and due to the polarisation factor $\mathbf{Q}\cdot \mathbf{e}$ in the one-phonon neutron scattering function, where $\mathbf{Q}$ is the wavevector and $\mathbf{e}$ is the mode eigenvector.
In the face-centred cubic (FCC) to tetragonal phase transition, crystal twinning can occur. In figure \ref{fig:phonons}(b) this has been accounted for by taking a weighted average of the three possible twins; the weights were adjusted empirically to get the greatest similarity between calculation and experiment. 

In figure \ref{fig:rot_dos}, both the low- and high-temperature phases were calculated using a supercell calculation followed by the band-unfolding method, outlined by \citet{Overy2017} (described below). For the high-temperature phase, this was necessary to take into account the disorder-induced phonon broadening; the low-temperature phase was calculated using the same method for consistency.

The band unfolding works as follows\citep{Overy2017}: the dynamical matrix of an $M_x\times M_y \times M_z$ supercell is diagonalised using GULP, resulting in $3ZN$ eigenvalues and corresponding eigenvectors, where $Z$ is the number of atoms in the unit cell and $N=M_x M_y M_z$ is the number of unit cells in the supercell. These $3ZN$ modes are then unfolded over the first Brillouin zone of the unit cell using the following equation:
\begin{align} \label{eq:projection}
    \rho(\mathbf{k},\omega)=\sum_\nu\left(\frac{\delta\left(\omega-\omega(\nu)\right)}{N}\sum_\alpha\sum_j\lvert\sum_{l}e_{jl}^\alpha(\nu)\exp{[i\mathbf{k}\cdot\mathbf{r}(jl)]}\rvert^2\right).
\end{align}
Here $\nu$ sums over the modes, $\alpha$ sums over Cartesian directions, $j$ sums over the unit cell atoms and $l$ sums over the unit cells in the supercell. $\omega$ is energy, $e_{jl}^\alpha(\nu)$ is the eigenvector for mode $\nu$ and unit cell atom $j$ in unit cell $l$ and $\mathbf{r}(jl)$ is the position vector for that atom. This equation holds for the allowed wavevectors $\mathbf{k}$ as determined by the supercell dimensions. In this work we use isotropic unit cells where $M_x=M_y=M_z=L$; the allowed wavevectors are:
\begin{align}
    \mathbf{k}_{\text{allowed}} = \left(\frac{n_{i}}{L}, \frac{n_{i}}{L}, \frac{n_{i}}{L} \right), \hfill n_i \in \{0, 1, ..., L-1\}.
\end{align}
Care must be taken when using a supercell of conventional rather than primitive unit cells. For the high-temperature phase, we use a supercell of $N$ conventional FCC unit cells. However, the sum over $l$ in eq. \ref{eq:projection} is over the \textit{primitive} unit cells, and there are additional allowed wavevectors in the supercell\citep{Boykin2006}, giving a total of $(4L)^3$ allowed wavevectors.

Eigenvector analysis was consequently performed with GASP\citep{Wells2002, Wells2015}, a software package that can, among other things, separate a molecular displacement into rotational and distortional components. For each eigenvector of the $\Gamma$-point calculation, the molecular structure was displaced along the eigenvector direction, with the average atomic displacement between 0.001 and 0.01 \AA. (The calculation was repeated with a displacement 10$\times$ larger, and yielded the same results. This means that the rotational/translational analysis is roughly independent of the displacement scale, at least for small displacements $<0.1$\AA.)
The structure was consequently relaxed in GASP, which here means that distortional displacements were reversed (since we assume that adamantane is rigid and arbitrarily high force constants were chosen, these distortional displacements are not meaningful). GASP's \texttt{netdr2} utility analyses, for each atom in each molecule, the total squared displacement:
\begin{align}
    M^2_{\text{tot, mole}} = \sum_i^{N_{\text{atoms}}} M^2_i
\end{align}
where $i$ sums over the atoms in the molecule and $M^2_i$ is the atomic displacement of atom $i$. Next, the \texttt{polycomp} utility analyses the squared displacement due to molecular rotation:
\begin{align}
    M^2_{\text{rot, mole}} = \sum_i^{N_{\text{atoms}}} M^2_{i, \text{rot}}
\end{align}
where $M^2_{i, \text{rot}}$ is the displacement of atom $i$ due to molecular rotation.
We define the rotational character as $M_{\text{rot}}/M_{\text{tot}}$.

\section*{Data availability}

Raw data from the neutron scattering experiments are available at \url{https://doi.org/10.5286/ISIS.E.RB1920588-1} and \url{https://doi.org/10.5286/ISIS.E.RB2010472-1}.

\section*{Acknowledgements}
Experiments at the ISIS Neutron and Muon Source were supported by a beamtime allocation RB1920588 and RB2010472 from the Science and Technology Facilities Council. HCW, AEP and RJCD thank EPSRC for financial support: EP/S035923/1 and EP/S03577X/1. BEM thanks ISIS for PhD studentship funding. The authors are indebted to R. Fair and M. D. Le for their assistance and guidance in the use of Euphonic, and to C. Goodway and M. Kibble for supporting the high pressure experiments. We are grateful to the UK Materials and Molecular Modelling Hub for computational resources, which is partially funded by EPSRC (EP/P020194/1 and EP/T022213/1).

\bibliographystyle{angew}
\providecommand*{\mcitethebibliography}{\thebibliography}
\csname @ifundefined\endcsname{endmcitethebibliography}
{\let\endmcitethebibliography\endthebibliography}{}

\end{document}